\documentstyle[aps,epsfig,multicol,subfigure]{revtex}
\begin{document}
\draft


\title{Relativity accommodates superluminal mean velocities}

\author{B. All\'es}

\address{INFN Sezione di Pisa, Pisa, Italy}

\maketitle

\begin{abstract}
Contrary to a widespread belief, measures of velocity can yield a value larger than $c$, the instantaneous light speed in vacuum,
without contradicting Einstein's relativity. Nevertheless, the effect turns out to be too small to explain the recently claimed
superluminal velocity by the OPERA collaboration. Several other general relativistic effects acting on the OPERA neutrinos
are also analyzed. All of them are unable to explain the OPERA result.
\end{abstract}

\pacs{04.20.-q; 04.20.Cv; 04.90.+e; 14.60.Lm}
\begin{multicols}{2}

\def\spose#1{\hbox to 0pt{#1\hss}}
\def\gtapprox{\mathrel{\spose{\lower 3pt\hbox{$\mathchar''218$}}
   \raise 2.0pt \hbox{$\mathchar''13E$}}}
\def\guu{$\geq$}

The recent observation of superluminal neutrinos by the OPERA team~\cite{adam} (see also the MINOS collaboration~\cite{adamson})
has prompted an intense discussion. Particle physics arguments~\cite{cohen,bi} or astrophysics data~\cite{gonzalez,fargion} seem to indicate
that neutrinos could hardly attain such velocities. Moreover, and generally speaking, it is accepted that
within the present physics knowledge, such velocities cannot be contemplated if the validity of the basic laws of relativity is to be
maintained. In particular the causal structure of spacetime would be badly violated if superluminal particles existed.

However it should not be forgotten that such a causal structure holds good in special relativity, namely, in an ideal
world where gravity is absent. When instead gravity is taken into account, a special relativistic spacetime arises only locally,
in a close neighbourhood of every spacetime event $P$. Strictly speaking, with infinite mathematial precision, this
is true exactly at $P$ (although, considering the inevitable inaccuracy of measurement devices, the region of validity of such
an approximation turns out to be more or less extended, depending on the magnitude of the spacetime curvature at $P$). Thus, it should
not come as a surprise that measurements carried out in a non--local way may apparently violate the main tenets of special
relativity.

We are going to show that, although the {\it instantaneous velocity} of a particle cannot be larger than $c$, relativity consents
that the {\it mean velocity} of the same particle may well exceed $c$. The interesting and appealing aspect of the above
statement is that what the OPERA collaboration has measured is precisely the neutrino mean velocity.

To illustrate the above effect it will be assumed that the spacetime is satisfactorily described by the metric
$ds^2=g_{00}dt^2-g_{ij}dx^idx^j$ where $(x^1,x^2,x^3)$ are general spatial coordinates and $t$ is the coordinate
time. Throughout the paper mass and time units will be chosen such that $c$ and the Newton constant $G$ are~1. Also, the
Einstein summation convention for pairs of repeated indices will be used with Latin indices indicating spatial components
and Greek indices referring indistinctly to both spatial and time components. It will be also supposed that all metric components
are time independent, $\partial g_{\mu\nu}/\partial t=0$, and that the time--space ones vanish, $g_{0i}=0$. This last condition is
not essential (we include it to avoid lengthy mathematical expressions) and can be omitted without changing the conclusions
of the paper~\cite{explanationsynchro}.

Consider a massless (or nearly massless) particle travelling along a spatial trajectory parametrized by a variable $\lambda$
and described by the three functions $(x^1(\lambda),x^2(\lambda),x^3(\lambda))$. It travels from the spatial point $P_1$ (where the parameter
takes the value $\lambda_1$) to the spatial point $P_2$ (value $\lambda_2$). After having measured the physical distance
$\Delta\ell$ between the two points (the metric is time independent), an observer at $P_2$ registers the time $\Delta\tau$
used by the particle to complete the trip. By physical distance we understand what one obtains for instance by lying rods in succession
between $P_1$ and $P_2$ (a more realistic procedure to extract $\Delta\ell$ will be described in the discussion of the OPERA experiment).
The time reading requires a previous synchronization among the clocks at $P_1$ and $P_2$. Then, the observer at $P_2$
defines the mean velocity of the particle as
\begin{equation}
\frac{\overline{v}}{c}\equiv\frac{\Delta\ell}{\Delta\tau}\;.
\label{0}
\end{equation}
Let us see what general relativity predicts for this quantity.

The physical distance $\Delta\ell$ is given by
\begin{equation}
\Delta\ell=\int_{\lambda_1}^{\lambda_2}d\lambda\,\sqrt{g_{ij}\dot{x}^i\dot{x}^j}\;,
\label{1}
\end{equation}
where $\dot{x}^i\equiv dx^i/d\lambda$. Viewed as a quadratic form, $g_{\ij}$ is positive definite.
On the other hand, one of the several equations that govern the motion of the particle
stems from the nullification of its proper time, $ds^2=0=g_{00}dt^2-g_{ij}dx^idx^j$, whence
\begin{equation}
\Delta t=\int_{\lambda_1}^{\lambda_2}d\lambda\,\frac{\sqrt{g_{ij}\dot{x}^i\dot{x}^j}}{\sqrt{g_{00}}}\;.
\label{2}
\end{equation}
Expression (\ref{2}) gives the coordinate time interval employed by the particle during the trip from $P_1$ to $P_2$. To
recover the time reading $\Delta\tau$ on the observer's clock at $P_2$ one has to multiply $\Delta\tau=\sqrt{g_{00}(\lambda_2)}\Delta t$.
Therefore the mean velocity recorded by the observer is
\begin{equation}
\frac{\overline{v}}{c}=\frac{\Delta\ell}{\Delta\tau}=\frac{1}{\sqrt{g_{00}(\lambda_2)}}\frac{\int_{\lambda_1}^{\lambda_2}d\lambda\,\sqrt{g_{ij}\dot{x}^i\dot{x}^j}}
{\int_{\lambda_1}^{\lambda_2}d\lambda\,\sqrt{g_{ij}\dot{x}^i\dot{x}^j/g_{00}}}\;.
\label{3}
\end{equation}
The dependence of $g_{\mu\nu}$ in (\ref{1})--(\ref{3}) on $\lambda$ comes through their dependence on the spatial coordinates.

The relevant fact that we wish to emphasize is that the mathematical framework of general relativity does not contain any mechanism
whatsoever able to oblige (\ref{3}) to stay equal or less than~$1$. Rather, as remarked above, (\ref{3}) tends to~1 only when the
velocity measurement becomes local, to wit, when points $P_1$ and $P_2$ tend to coincide. Indeed, since $\sqrt{g_{ij}\dot{x}^i\dot{x}^j}$ is
positive, the mean value theorem can be applied and it states that a value $\lambda_\xi$, comprised between $\lambda_1$ and
$\lambda_2$, exists such that $\int_{\lambda_1}^{\lambda_2}d\lambda\,\sqrt{g_{ij}\dot{x}^i\dot{x}^j/g_{00}}=(1/\sqrt{g_{00}(\lambda_\xi)})
\int_{\lambda_1}^{\lambda_2}d\lambda\,\sqrt{g_{ij}\dot{x}^i\dot{x}^j}$. Inserting this expression in the denominator of (\ref{3}) we obtain
\begin{equation}
\frac{\overline{v}}{c}=\frac{\sqrt{g_{00}(\lambda_\xi)}}{\sqrt{g_{00}(\lambda_2)}}\;,
\label{4}
\end{equation}
which tends to unity as $P_1\to P_2$ because in this limit $\lambda_\xi\to\lambda_2$
also.

Let us consider an instance of the effect just exposed by studying the radial motion of a particle in a Schwarzschild metric
(it characterizes the vacuum outside a spherically symmetric mass distribution with total mass $m$) in standard coordinates,
$ds^2=g_{00}dt^2-g_{rr}dr^2-r^2d\Omega^2$ where $d\Omega$ is the
solid angle and $g_{00}=1/g_{rr}=1-2m/r$. The massless or almost massless particle will travel radially from $r_1$ to $r_2$
(we do not specify yet whether $r_1<r_2$ or $r_2<r_1$). Because $g_{00}$ increases with $r$, following (\ref{4}), we deduce $\overline{v}<c$
if $r_2>r_1$ and $\overline{v}>c$ if $r_2<r_1$. The detachment from the light speed can be better appreciated by plotting $\overline{v}/c$
against $r_1$ and $r_2$. Specifically, for the Schwarzschild metric we have (calling $\rho_{_A}\equiv r_{_A}/(2m)$, $A=1,2$)
\begin{eqnarray}
\frac{\Delta\ell}{2m}&=&\Bigg\vert\sqrt{\rho_2^2-\rho_2}-\sqrt{\rho_1^2-\rho_1}\nonumber\\
                      &&+\frac{1}{2}\log\frac
{(\sqrt{1-1/\rho_2}+1)(\sqrt{1-1/\rho_1}-1)}{(\sqrt{1-1/\rho_2}-1)(\sqrt{1-1/\rho_1}+1)}\Bigg\vert\;,\nonumber\\
\frac{\Delta\tau}{2m}&=&\sqrt{1-1/\rho_2}\;\left\vert\rho_2-\rho_1+\log\frac{\rho_2-1}{\rho_1-1}\right\vert\;.
\label{5}
\end{eqnarray}
The absolute values in (\ref{5}) make these expressions valid for both cases, $r_2>r_1$ and $r_2<r_1$. This is an important remark because
the fact that the time intervals for the particle to go from $r_1$ to $r_2$ or for coming from $r_2$ to $r_1$ are the same allows to reliably
synchronize clocks with the exchange of light signals.

In Figs.~1 and~2 the mean velocity $\overline{v}/c$ is plotted against $1/\rho_1$ for various values of $\rho_2$. Note {\it (i)} that
we have deliberately excluded the region $1/\rho_1>1$ because otherwise the particle would enter the horizon of the corresponding black hole
and {\it (ii)} that all plots touch the line $\overline{v}/c=1$ when $\rho_1$ coincides with $\rho_2$, confirming the validity of special relativity at short
distances.

The case $r_1>r_2$ (the particle approaching the spherical distribution of mass) is shown in Fig~1. As viewed by the observer in $r_2$,
$\overline{v}$ turns out to be always larger than $c$. This effect is particularly pronounced when the observer is close to the
Schwarzschild radius $r_s\equiv2m$ (lower values of $\rho_2$) and for large separations $r_1-r_2$. For $\rho_2>100$ the
resulting $\overline{v}/c$ is so close to~1 that the related plots cannot be seen within the scale of the vertical axis.

In Fig.~2 the case $r_1<r_2$ (the particle going away from the spherical massive object) is treated. Now all mean velocities come smaller
than $c$, which looks quite surprising for massless particles. Again $\overline{v}$ tends to $c$ whenever the radial coordinate $r_1$ of
the position of the particle's departure is close to the radial coordinate $r_2$ of the position of arrival.
Note the tendency of $\overline{v}$ to become null as $r_1$ approaches $r_s$ for every $r_2$.

\vskip 5mm

\begin{figure}
\includegraphics[scale=0.33]{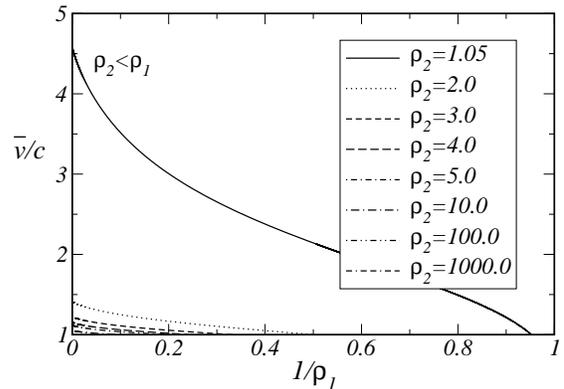}
\caption{Mean velocity $\overline{v}/c$ of the particle travelling from $r_1=2m\rho_1$ to $r_2=2m\rho_2$ in a Schwarzschild metric for
several values of $\rho_2$ as a function of $1/\rho_1$ and always with $r_1>r_2$.}
\label{Fig1}
\end{figure}

\vskip 3mm

\begin{figure}
\includegraphics[scale=0.33]{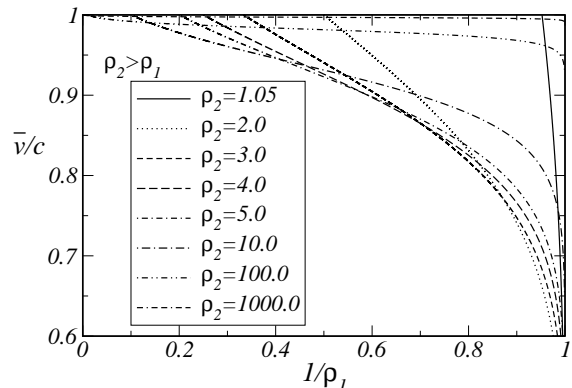}
\caption{Mean velocity $\overline{v}/c$ of the particle travelling from $r_1=2m\rho_1$ to $r_2=2m\rho_2$  in a Schwarzschild metric for
several values of $\rho_2$ as a function of $1/\rho_1$ and always with $r_2>r_1$.}
\label{Fig2}
\end{figure}

As shown in the above plots, the superluminality or subluminality of the particle's mean velocity is dramatically enhanced in regions
of large curvature (near the Schwarzschild radius $r_s$). This stresses the general relativistic character of the the described effect.

A terrestrial experiment prepared to detect values for the radial velocity $\overline{v}$ different from $c$ cannot approach any
Schwarzschild radius. If such an experimental set--up was constrained to use $r_1,r_2\geq r_\oplus$, the Earth's radius, 
then $\Delta\ell$ and $\Delta\tau$ in (\ref{5}) could be approximated with excellent accuracy to linear order in $m_\oplus/r_{_A}$ ($m_\oplus$
is the Earth mass) and (\ref{0}) would barely differ from~1,
\begin{equation}
\frac{\overline{v}}{c}\approx1+\frac{m_\oplus}{r_\oplus}-\frac{m_\oplus}{r_1-r_\oplus}\log\frac{r_1}{r_\oplus}>1\;,
\label{6}
\end{equation}
where the observer has been put on the Earth's surface, $r_2=r_\oplus$, and we have taken $r_1>r_\oplus$.
For $r_1$ not much larger than $r_\oplus$, the percentage of excess of velocity is a meagre $(\overline{v}-c)/c\sim10^{-10}$.

To exemplify the above findings, we apply them in two cases: first, to the already mentioned OPERA experiment and,
second, to the determination of the distance from the Earth to the Moon by exchange of light pulses.

In the OPERA experiment a beam of muon neutrinos was produced at CERN SPS and sent to the Gran Sasso underground laboratories
(LNGS) in Italy where they were revealed~\cite{adam}. The baseline distance $\Delta\ell$ was counted from the place at CERN where
the proton beam time--structure is being ascertained (it is the so-called Beam Current Transformer or BCT) to the origin of the OPERA
detector at LNGS. The coordinates in
the Universal reference frame ETRF2000 \cite{boucher} of the two locations were obtained by two steps: firstly the coordinates of
ancillary benchmarks placed outside the two laboratories were determined by a GPS campaign and secondly the distance between these
benchmarks and the BCT at CERN or the OPERA detector at LNGS was directly measured by geodetic survey. The value of $\Delta\ell$ was then derived from the
coordinates by usual Euclidean geometry \cite{colosimo} yielding $731278.0\pm0.2$ m (deriving it from the spatial part of the metric as in (\ref{1}) would
add a neglibible correction). Effects from geoid undulation, crust tides and continental drift (including earthquakes) do not significantly
modify the above result. After an accurate synchronization between clocks of both laboratories~\cite{tedesc},
the baseline length was divided by the time of flight of neutrinos to obtain their average velocity. The result was larger than $c$ by an
amount $[(\overline{v}-c)/c]_{\rm experiment}=2.48\pm0.28({\rm stat})\pm0.30({\rm syst})\cdot10^{-5}$. However, the mechanism described in the paper is
unable to explain this excess. This can be easily seen by adopting the simplifying hypothesis that neutrinos travel over the sphere of the Earth's surface
and taking advantage of the fact that the angular sector of the Schwarzschild metric in standard coordinates is flat. Then $\Delta\ell$ equals
the Euclidean result $r_\oplus\Delta\theta$ ($\Delta\theta$ is the angular separation between CERN and LNGS~\cite{obtain}) while
$\Delta t$ is $r_\oplus\Delta\theta/\sqrt{1-2m_\oplus/r_\oplus}$. Hence $\overline{v}/c$ is always~1. Eliminating
the previous simplifying hypothesis (neutrinos actually traversed the Earth's crust following the imaginary chord that joins CERN and LGNS)
adds a negligible contribution of opposite sign (to verify this assertion an interior metric was derived at first order in $m_\oplus$ assuming a
planet Earth with uniform mass density). The inclusion of the spin of Earth (for example by using the Lense--Thirring metric) produces an even
smaller contribution (while $m_\oplus/r_\oplus\sim7\cdot10^{-10}$, $J_\oplus/r_\oplus^2\sim4\cdot10^{-16}$, $J_\oplus$ being the Earth's angular momentum).
Only the rotation of the observer's laboratory at Gran Sasso induces a revealable effect but it affects the synchronization of clocks
and therefore it has nothing to do with the topic described in this paper~\cite{synchroCERN}.

These considerations seem to imply that the observation of a superluminal velocity for the neutrinos of OPERA must be likely ascribable to other
reasons: either purely experimental oversights~\cite{bergeron,dado} or really new physics peeping out
(see~\cite{amelino,tamburini,klinkhamer1,giudice,dvali,mann,drago,li,iorio,alexandre,nicolaidis,klinkhamer2,ciuffoli,anber,nojiri,mecozzi,morris1,morris2,pavsic,bramante,schreck}
for a partial list of theoretical suggestions and criticisms).

Consider now the lunar laser ranging experiment (LLR) \cite{dickey,williams} in which, among other ephemeris, the Earth--Moon distance is determined.
It consists in sending a laser pulse from the Earth to the lunar surface where it is reflected (several manned missions in the past left corner reflectors
on the Moon's surface) and received back on Earth. Multiplying the time employed by the light in its round trip by the speed of light yields the
Earth--Moon distance. However, assuming that the metric is well--approximated by the Schwarzschild one at linear order in $m_\oplus/r_\oplus$,
the mean speed of the light ray turns out to be larger than $c$ and so, the distance should come out less than what it really is. To remedy this
inconvenient and obtain at least an order of magnitude of the necessary correction, we resort again to expressions (\ref{5}).

Establishing that the radial coordinate of the laboratory is the Earth's radius $r_\oplus$, it has yet to find the radial coordinate on
the Moon, $r_1$. This is achieved by inverting the formula for $\Delta\tau$ in (\ref{5}). After some algebra we get the equation
\begin{equation}
\exp\left[\left(\frac{\Delta\tau}{r_\oplus-m_\oplus}+1\right)\;\rho_\oplus\right]=\xi\;e^{\xi\rho_\oplus}\;,
\label{7}
\end{equation}
where $\xi\equiv r_1/r_\oplus$, $\rho_\oplus\equiv r_\oplus/(2m_\oplus)$ and $\Delta\tau$ is half of the time (measured by terrestrial clocks)
spent during the round trip. Eq.(\ref{7}) can be solved for $\xi$ in terms of the Lambert $W$--function~\cite{corless}. However, on
account of the fact that $\rho_\oplus\gg1$, we can approximately set $\xi\approx1+\Delta\tau/(r_\oplus-m_\oplus)$. Inserting
it into the first of (\ref{5}) leads to
\begin{equation}
\hbox{Earth--Moon distance}\approx\Delta\tau+\frac{2m_\oplus}{r_\oplus}\Delta\tau\;,
\label{8}
\end{equation}
the correction being $2m_\oplus\Delta\tau/r_\oplus\approx53$ cm. It must be stressed that the above exercise has been presented in order
to display an instance where the issue discussed in this paper yields a not insignificant contribution. With another name, this effect has surely
been taken into account by the LLR collaboration. Indeed, the time lapse during the round trip of the laser pulse was evaluated in barycentric
reference frame coordinates~\cite{williams,moyer} by including all general relativistic effects and leading to the Shapiro formula~\cite{shapiro}.
Moreover, in the original calculation of the LLR collaboration, the effects from all main Solar System bodies are included.

In conclusion we have seen that mean velocities in general do not conform to well--known special relativity principles. In particular, albeit amazing,
the average velocity at which a particle has travelled for a long time can be different from the instantaneous velocities that the particle
attained at every point along the trajectory, even when the latter was the same at all points. Indeed, special relativity
adequately describes physics only locally while the average over large distances of a velocity must necessarily be introduced as a non--local
quantity. Utilizing these considerations for understanding the results from the
OPERA collaboration, we conclude that the real anomaly is {\it not} that $(\overline{v}-c)/c$ be positive but that it is a rather large number.

The discussion about mean velocities can be straightforwardly generalized to time dependent metrics as sensible definitions of proper time
and length are also admissible on such metrics. An instance is the Friedmann--Robertson--Walker metric.

For the Schwarzschild metric, the effect described in the paper is driven mainly by the curvature dependence of the proper time $\Delta\tau$.
Indeed, note that as $\rho_1$ approaches~1, the value of $\Delta\tau$ diverges while that of $\Delta\ell$ stays finite for any values of $\rho_1$
and $\rho_2$. This property is not general. For example, in a conformally flat metric $g_{\mu\nu}=f(x)\eta_{\mu\nu}$,
$\Delta\tau=\sqrt{f(x_2)}\Delta t$ while $\Delta\ell=\int dx\sqrt{f(x)}$, which shows that any divergent behavior of the function $f(x)$ may influence
both $\Delta\tau$ and $\Delta\ell$. An example of a conformally flat metric is again the Friedmann--Robertson--Walker metric, although it is more
usually presented in Gaussian coordinates.

We have analyzed the case in which the observer stays at rest at one end of the particle's trajectory. But of course other experimental
dispositions are possible: the observer staying in the middle of the particle's trajectory or even at a point outside it;
the observer not at rest, etc. Also many possible definitions of mean velocity, other than the one used by OPERA (\ref{0}), can be
conceived. In all cases bizarre results should be carefully interpreted.

\vskip 5mm

It is a pleasure to thank Mihail Mintchev and Giancarlo Cella for stimulating discussions.

{\it Note added:} During the refereeing of the paper we became aware of the existence of Ref.~\cite{lust} where a similar
analysis is performed. However the study presented here is more general and detailed.

\end{multicols}

\begin{references}
\bibitem{explanationsynchro} If the metric contained non--zero time--space terms $g_{0i}$ then, the spatial distance in (\ref{1})
would be evaluated with the positive definite quadratic form $g_{0i}g_{0j}/g_{00}+g_{ij}$~\cite{moller} and the synchronization of coordinate
times at distant points would require the inclusion of a lag defined by $-\int_{\lambda_1}^{\lambda_2}d\lambda\,g_{0i}\dot{x}^i/g_{00}$.
Apart from these complications, the effect presented in the text would be, {\it mutatis mutandis}, qualitatively the same.
\bibitem{synchroCERN} An event occurring at coordinate time $t_{\rm CERN}$ at CERN is simultaneous with another event
occurring at Gran Sasso at coordinate time $t_{\rm LNGS}$ if $t_{\rm LNGS}-t_{\rm CERN}\approx-\omega_\oplus
(r_\oplus\cos\alpha)^2\Delta\varphi\approx-2.3$ ns ($\omega_\oplus$ is the angular speed of the Earth's rotation, $\alpha$ is
the latitude of the CERN or Gran Sasso ---assumed equal--- and $\Delta\varphi$ is the difference in longitude of both
laboratories). This time lag is related to the Sagnac effect and it has been discussed in~\cite{colosimo}.
$-2.3\pm0.9$ ns is precisely the systematic error identified in~\cite{tedesc}.
\bibitem{obtain} $\Delta\theta$ is obtainable from the geographical coordinates of Geneva (latitude=46$^{\rm o}$ $12^\prime$,
longitude=$6^{\rm o}$ $9^\prime$) and Gran Sasso (latitude=42$^{\rm o}$ $28^\prime$, longitude=$13^{\rm o}$ $33^\prime$).
\end{references}
\end{document}